\documentclass[journal,12pt,]{IEEEtran}
\usepackage{cite}
  \usepackage[pdftex]{graphicx}
  \graphicspath{{./}}
  \DeclareGraphicsExtensions{.pdf,.jpeg,.png}
\usepackage{amsmath}
\usepackage{algorithmic}
\usepackage{array}
\usepackage[inline]{enumitem}
\usepackage{stfloats}
  \usepackage[caption=false,font=footnotesize]{subfig}
\usepackage{url}
\usepackage[dvipsnames]{xcolor}

\begin{document}
\title{Service-based Fog architecture without DNS redirection}

\author{Mays~AL-Naday,~\IEEEmembership{Member,~IEEE,}
        Martin~J.~Reed,~\IEEEmembership{Member,~IEEE,}
        Janne~Riihij\"arvi,~\IEEEmembership{Member,~IEEE,}
        Dirk~Trossen,~\IEEEmembership{Member,~IEEE,}
        Nikolaos~Thomos,~\IEEEmembership{Senior Member,~IEEE,}
        and~Mohammed~Al-Khalidi,~\IEEEmembership{Senior Member,~IEEE}% <-this % stops a space
}

\maketitle

\begin{abstract}
The heterogeneous and distributed nature of the Internet of Things (IoT) is driving the need for extremely fast and fine-grained service provisioning in 5/5+G architectures and beyond. To meet these needs, it is critical to enable efficient and flexible computation and networking fabrics, that can be rapidly reconfigured to meet the computation and communication tasks at hand. In this article, we propose a novel Fog Computing architecture that translates IoT communications into service transactions, provisioned over a fast and efficient networking fabric. Service matching is provided by a network function designed using principles from Information-Centric Networks (ICN) research, routing edge requests directly to the nearest service points without expensive and slow DNS redirects. The proposed Fog substrate reduces the networking complexity and overhead while being architecturally simple. We evaluate the architecture through a comparison with how Fog might be established over existing networking fabrics and quantify the performance benefits. Evaluation results illustrate the superiority of the proposed architecture in reducing the required backhaul capacity and the path length.
\end{abstract}

\begin{IEEEkeywords}
IoT, Fog Computing, Information-Centric Networking, Virtual Network Function
\end{IEEEkeywords}

% \fi
\IEEEpeerreviewmaketitle

\section*{Introduction}
\IEEEPARstart{F}{og} computing has emerged as an extension of cloud computing at the network edge~\cite{Mouradian2018}. It is envisaged as a highly virtualized platform that offers compute, storage and networking capabilities closer to the end-users; thereby, localising and personalising many of the offered services. Fog computing supports a variety of new services, such as personalised content delivery in 5/5+G~\cite{Xiaoqing2015}; and, networking 'smart devices' (\emph{things}) \cite{OpenFog2017}. The latter encompasses a rapidly growing range of Internet of Things (IoT) scenarios, in which end-user devices can communicate with the \emph{Fog} to generate, retrieve and/or process information, allowing for fine-grained (and increasingly automated) management and instrumentation of end-user environments. This form of communications is naturally highly distributed, heterogeneous and possibly sensitive to latency and privacy constraints. The latter, in particular, provides motivation for localised computation and networking in the Fog \cite{Xiaoqing2015,OpenFog2017}.

To realise the objectives of Fog computing, it is critical to define the interplay between the Fog and the Cloud, and to have efficient communications among Fog nodes, and between Fog and Cloud nodes. Cisco and the OpenFog consortium defines this interplay as a 'multi-tier' architecture of both cloud and Fog nodes \cite{Xiaoqing2015,OpenFog2017}. In this architecture, nodes at the edge (i.e lowest tier) are of constrained capacities (compute/storage/network) but they provide a response within a short time period; thereby, they support data collection and process offloading. In contrast, nodes in the cloud core (i.e highest tier) have much lower constraints on their capacities and much higher time-to-response. Consequently, the cloud core generally supports deep data processing and analysis to generate global knowledge. In between, there could be a number of intermediary tiers, responsible for aggregating data from lower tiers and generating knowledge that feeds to the cloud.

From a network perspective, the Fog architecture has similarities to that of Content Distribution Networks (CDNs) \cite{Xiaoqing2015}; exhibiting similar challenges when it comes to facilitating the communication between different Fog nodes, albeit at a considerably larger scale \cite{Xiaoqing2015}. So far, Cloud and CDN solutions have been facilitated through traditional network architectures. The latter, strongly tie information to location, namely in the Domain Naming Service (DNS) and the routing/forwarding functions. This makes the network intrinsically reliant on such mechanisms as \emph{DNS-redirection} to facilitate \emph{anycast} communications. Moreover, current research in Fog computing, generally assumes the use of such a traditional network architecture \cite{Cirani2015}. However, experience with CDNs has shown that existing DNS redirection mechanisms are highly rigid and inefficient approaches~\cite{Sitaraman2014}; resulting in ongoing research to mitigate this sub-optimal approach~\cite{Chen2015}.

DNS redirection provides a mechanism to maintain multiple replica servers with unique IP addresses for a single fully qualified domain name (FQDN) such that the FQDN can be ``mapped'' to one of the server IP addresses, depending on the location of the client. DNS redirection is, currently, the most cost effective form of such redirection, as it only depends on the DNS server location and incurs much lower overhead than other forms of redirection, such as HTTP redirection, which requires deep-packet inspection. However, DNS redirection relies on selecting a ``nearest'' replica to the DNS point, not necessarily to the client.
%based on the client's choice of DNS server without any accurate topological information. 
This results in some scenarios having highly inefficient mappings; thereby, degrading the time-to-response while increasing load inside the network. The solution of~\cite{Chen2015} mitigates this inefficiency by extending traditional DNS with information about end-user subnets; allowing for mapping a subnet to the nearest replica. To maintain a manageable DNS state, a general agreement on a subnet mask length of $/20$ has been assumed~\cite{Chen2015}, which translates into a subnet of up to $4094$ users. While this has been a reasonably small subnet in existing CDNs; the size and distribution of Fog substrates is expected to demand significantly higher granularity and much less complexity.

This paper proposes a novel Fog architecture that translates Fog-to-Fog and Fog-to-Cloud interactions to managed service transactions, identified by pre-defined wild-card URLs. In this context, Fog and Cloud nodes act either as consumption points or service points, depending on whether they request or offer services. Matching supply with demand is provided through a novel service routing solution, embraced from emerging research in Information-Centric Networking (ICN); and, directly realized over a L2 transport network~\cite{Trossen2015}. Such \emph{service routing} eliminates the need for DNS and HTTP redirection, whilst utilizing an efficient, agile, L2 multicast solution for transport in the network and therefore removing the need for any form of application layer multicast.

% Matching supply with demand is provided without DNS redirection and content reflection from CDN, while allowing for flexible, and highly agile, composition of anycast/multicast relationships; therefore, aiming to positively impact user-level latency as well as network utilization. 
% This is achieved by interpreting user requests to CDNs, as well as the transport from the origin server to CDNs, as a service transaction. We employ a novel service routing solution, directly realized over a L2 transport network. Such service routing eliminates the need for DNS and HTTP redirection, whilst utilizing an efficient, agile, L2 multicast solution for transport in the network and therefore removing the need for reflectors. In our evaluation, we focus on the efficiency gains of our proposed mapping%, achieved by using our ICN routing function
% %removing the need for DNS-based indirection
% , leaving the improvements of the transport for future work. 
%  For our realization, we utilize insights from the emerging research in Information-Centric Networking (ICN) that natively supports \emph{anycast}, while path calculation and forwarding is facilitated through source-routing mechanisms that natively support multicast in the network.
 Whilst the proposed realization is rooted in first deployments of real prototypes in early trials, it is supported by first contributions to standardization, most prominently in the IETF through Service Function Chaining (SFC)~\cite{ietf2017}. An early realization of the service routing substrate has also been part of an ETSI Multi-access Edge Computing (MEC) proof-of-concept~\cite{mec2017},
which has showcased efficient content delivery for HTTP-level streaming as well as edge content retrieval in localized applications.

Our contributions in this paper are twofold:
\begin{enumerate*}
\item First, we describe the design of a novel, service-based, Fog architecture, relieved from DNS redirection; and, instead, comprising a set of functions focused on providing service management, routing and resource management. We present the consumption and service points as edge clusters that communicate over a fast and agile Point-2-Multi-Point (P2MP) core. We describe the network function utilized in the core to provide edge-to-edge service routing and show how it can efficiently replace DNS redirection to provide \emph{anycast}. Moreover, with the proposed  source-routing mechanism, we show how quasi-synchronous unicast requests at the edge can be spontaneously grouped into a multicast stream and be delivered to the service point as a single unicast request; to which, the service point would issue a single response. This would not only optimise the utilization of network resources, but also allows a service point to drastically reduce transmitted bits per request; thereby, satisfying larger demands with an equivalent server load.  
% \item Second, we define the model to describe our proposed architecture and formulate the resource placement problem as a variance of the ``$K$-center'' problem.
% %\item Second, we define the mathematical model to describe the relationships between publications and subscriptions, and the resultant traffic in the fCDN. We use this model to define the resource requirements in the network, in terms of storage and link capacities; and formulate the resource placement problem as a variance of the ``$K$-center'' problem. 
\item Second, we analyse the performance of the proposed substrate and compare its performance to that of currently proposed, DNS-based, Fog architectures. We quantify the benefits from eliminating the suboptimal point selection of DNS, and the benefits of spontaneous multicast in the network.
\end{enumerate*}

In the rest of this paper we first presents the proposed Fog substrate, describing the network functions and the service routing required at the edge. We then introduce the problem of service placement, showing its impact on the user's perceived experience of latency. The service placement problem is solved through pseudo random selection algorithms that are evaluated in the proposed Fog architecture; showing it has significant benefits over traditional, DNS-redirection based, architectures.

\section*{Proposed Fog Architecture}
%\label{sec:fog_arch}
%-------------------------------------------
Before we introduce the Fog architecture, we briefly introduce the foundational work on realizing an IP routing substrate on top of an ICN solution, in turn realized over an L2 transport network. From this original work, we formulate the proposal for a flexible, service-based Fog architecture. We focus on web services supported by the HTTP and CoAP protocols as they dominate the demand of end-users; however, the solution is applicable to any chunk-based protocol. We position the solution within the scope of a local Fog substrate in an operator's network; and assume wider connectivity to be realised through an inter-connected set of Fog clusters.

% will expand into our Fog architecture. We 
%-------------------------------------------
Mapping IP-based protocols, such as HTTP and COAP, into ICN abstractions has been proposed in~\cite{Trossen2015}. The approach comes as a feasible migration story towards ICN, which significantly reduces the required upgrades to existing infrastructure. This is achieved by limiting change to the core of an operator's network; whilst exploiting (and integrating with) rapidly developing technologies such as Software-Defined Networking (SDN) and Network Function Virtualisation (NFV). This has attracted the attention of stakeholders and now is moving towards standardization~\cite{ietf2017,mec2017}. 

The IP-over-ICN architecture divides an operator's network into two parts: a set of IP networks at the edge; and, a Point-to-Multi-Point (P2MP) forwarding core that connects the edge nodes/networks. The core provides line-speed switching, and a set of network functions that facilitate matching supply with demand and creating edge-to-edge paths. Those functions are: Rendezvous (RV), for name matching, and a Topology Management (TM) to provide routing~\cite{Fotiou2012}. Forwarding in the core is enabled through a highly efficient forwarding (FW) function, such as that proposed by~\cite{Jokela2009} and more recently its variant suitable for direct SDN deployment~\cite{Reed2016}.
An IP edge is connected to the core through a Service Router (SR) that provides mapping of traditional IP-based protocols (IP, HTTP, etc.) to, and from, a namespace managed by the RV function. Furthermore, the SR  translates the request/receive semantics into an equivalent publish/subscribe counterpart~\cite{Trossen2015}. In this context, a server listening for requests of a service is represented by an ICN \emph{subscription} for requests; while a request for a service is translated into an ICN publication. The Pub/Sub roles are reversed in the response direction.

The architecture above loosens the tie between information and location, as it treats IP-based protocols as services, decoupled from the location of their actuators. The novel Fog substrate utilises the capabilities of this architecture to provide efficient anycast and multicast service routing and extends the granularity of service matching beyond protocol level, into \emph{micro-services} based on wild card URI matching. 
Below, we present the proposal for a service-based Fog architecture. We show that by using the ICN matching of supply with demand, we no longer need DNS redirection or DNS extensions to provide the anycast mapping. Furthermore, the network is natively supporting multicast, thereby eliminating the need for application-layer multicast.
\subsection*{Fog: Micro Services}
%\label{sec:fog_over_icn}
Based on the initial ideas of IP-over-ICN, the key aspect in the proposed architecture is the extension of service granularity beyond the mere protocol level. Notably, while the base architecture maps a FQDN to a \emph{service}, here, we extend this mapping to individual, managed, resources of a FQDN. Such resources may be chunks of content, data and/or a set of function(s), normally aggregated under wild-card URLs, ``stream identifiers'' or proprietary URIs. We map such identifiers to \emph{micro-services}, encompassed within the FQDN service. Following this logic, we interpret Fog/Cloud nodes as both consumption and service points, depending on whether they are requesting or offering services/micro-services. User-to-Fog traffic is constrained locally; while, Fog-to-Fog and Fog-to-Cloud requests (and responses) are translated into micro-service transactions, routed over the network entirely at the relevant HTTP (or COAP) service level. 

\begin{figure}[tb]
\includegraphics[width=1.0\linewidth]{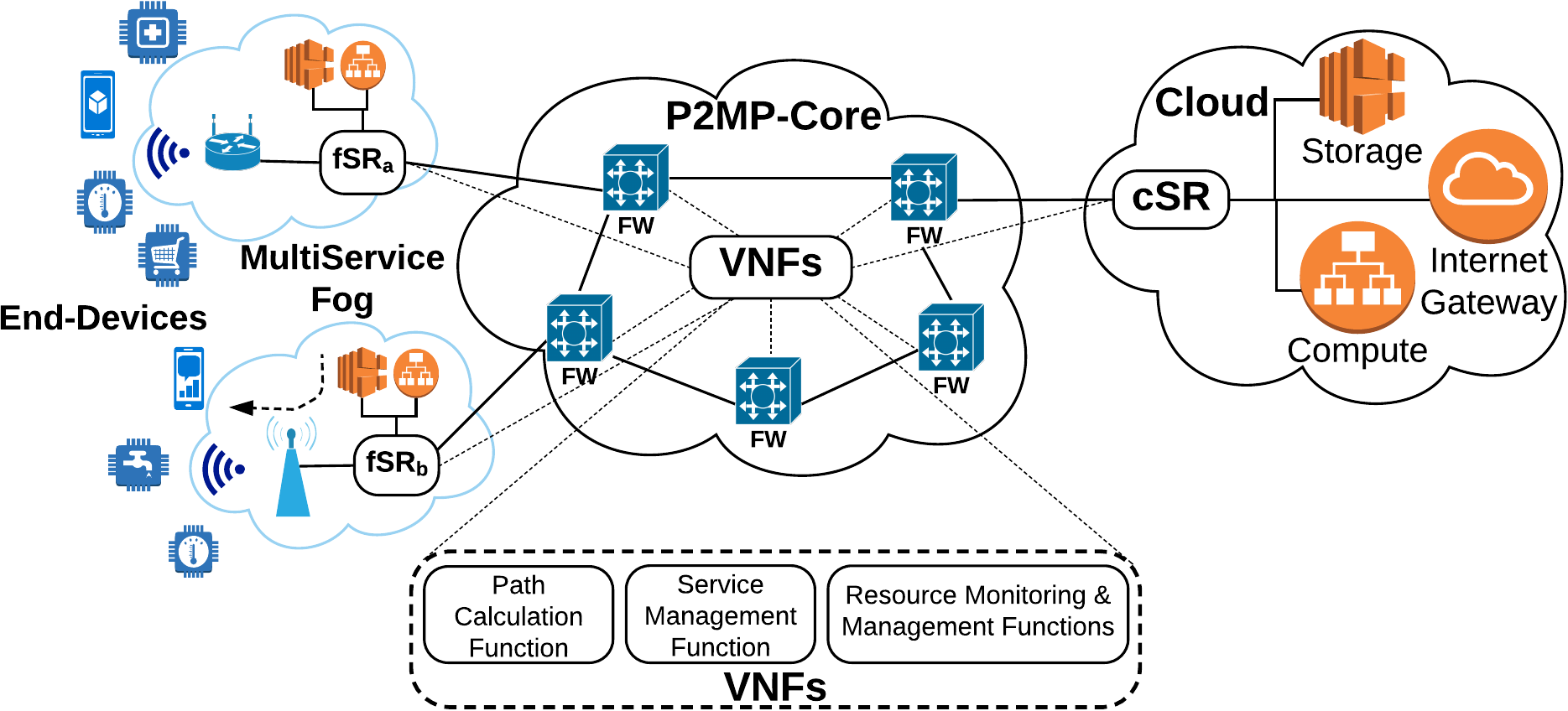}
\caption{A functional view of the proposed Fog architecture, showing the Virtual Network Functions (VNFs) managing the network, including a Path Computation Function (PCF), responsible for Pub/Sub matching and creating source-routed multicast trees between service and consumption points. Other functions include: a Service Management Function (SMF) and a Resource Monitoring and Management Function (RMMF)}
\label{fig:fog_arch}
\end{figure}

\begin{figure}[tb]
	\centering
	\subfloat [Name-Space View (HTTP/Micro-HTTP)]{
				\includegraphics[width=1\linewidth]{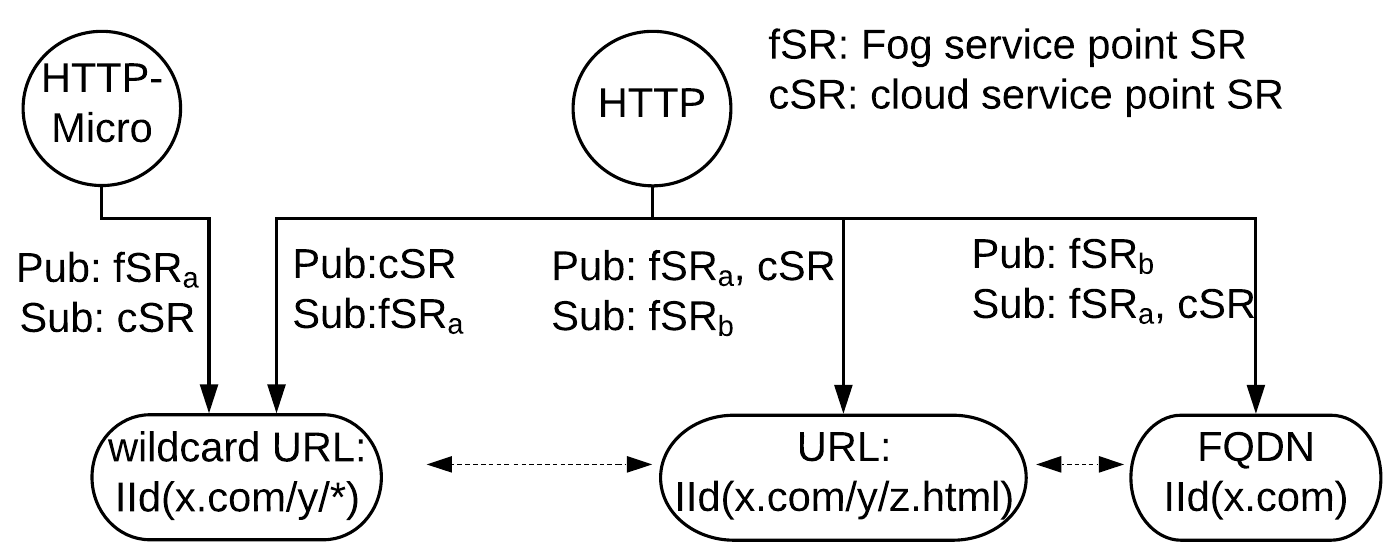}
	}
		\label{fig:fog_ns}\vfill 
	\subfloat [Sequence of Publications]
	{
				\includegraphics[width=1\linewidth]{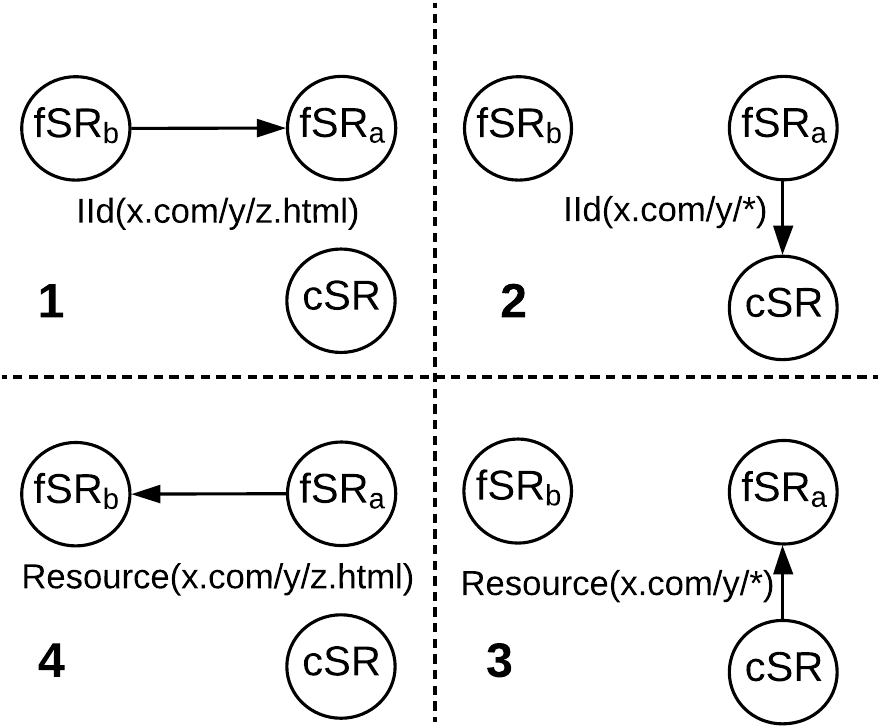}
	}
		\label{fig:fog_ms}
        \\\vspace{0em}
	\caption{The proposed Fog architecture viewed as a name-space and sequence of publication/subscription events. The name-space view, (a), illustrates the naming hierarchy of the service and its related micro-service, as well as the published/subscribed IIds by the SRs in the example network of Figure~\ref{fig:fog_arch}. The publications view, (b), illustrates the sequence of publications incurred when SR$_b$ requests a none-cached resource}
    \label{fig:fog}
\end{figure}
The proposed architecture is depicted in the example network of Figure~\ref{fig:fog_arch}. A network operator deploys a fast-switching, Point-to-Multi-Point (P2MP) substrate in the core; capable of performing stateless-multicast forwarding. Such a fabric may be facilitated with the Bloom Filter-based solution of~\cite{Jokela2009}, its Bit-based variance in SDN~\cite{Reed2016}, or the Bit Indexed Explicit Replication (BIER) solution~\cite{Giorgetti2017}. For the rest of this work, we will assume the solution of~\cite{Reed2016} to be the forwarding mechanism in the network core. To manage and facilitate service/micro-service transactions, the core introduces a set of Virtual Network Functions (VNF)s: \emph{Path Calculation Function (PCF)}, \emph{Service Management Function (SMF)} and \emph{Resource Management and Monitoring Function (RMMF)}. The PCF matches service requests (i.e. ICN publications) with service offerings (i.e. ICN subscriptions), resulting in a suitable path calculation as described in~\cite{Fotiou2012,Trossen2015}. Notably, with the PCF mapping consumption points to service points, there is no longer a need to have mapping through DNS-redirection. This alleviates the network from the overhead of DNS-redirection and allows for optimized mapping to the true ``best'' service point. The SMF manages the placement of service points, and the distribution of services across points; whilst the RMMF manages the compute, storage and network resources.

Outside the core network, multiple Fog clusters are present. Each cluster may connect multiple end-users and/or smart end-devices (i.e.\emph{things}); as well as a Fog Node to provide compute and storage resources (constrained or otherwise). At the termination point of a cluster, a Fog SR (fSR) is presented to connect the cluster to the core and to other Fog/Cloud clusters. The fSR is responsible for: subscribing to the list of services/micro-services offered by the Fog node, mapping requests/responses of the Fog node to the appropriate service name; and, publishing on behalf of the Fog node. A Cloud cluster (or multiple clusters) is also connected to the core via a cloud SR (cSR), which publishes/subscribes on behalf of the Cloud nodes in a similar manner to the fSR. The difference between Fog and Cloud nodes are that the latter would not connect end-users or things, only compute and storage resources; as well as linking the network to other networks and the Internet. 

% When the Fog node is requesting a service, the SR acts as a consumption point router (uSR) and publishes the request for a service . In contrast, when the Fog node is offering a service (i.e. listening to requests for the service) the SR acts as a Fog service point router (fSR) and subscriber to requests for the service.  A fSR may also publish a request for a none-cached micro-service, as A Cloud cluster (or multiple) is also connected to the core via a cloud SR (cSR), which publish/subscribe on behalf of the cloud nodes in a similar procedure to the f/uSR. The difference between Fog and Cloud nodes is that the latter would not connect end-users or things, only compute and storage resources; as well as linking network to the Internet. 

\subsection*{Managed Service Naming}
The name-spaces to be supported in the proposed Fog architecture may vary for different services. Here we describe an example HTTP name-space, illustrated in Figure~\ref{fig:fog}(a), focusing on web services. We show through Figure~\ref{fig:fog}(a) how the proposed Fog architecture defines names for services and their respective Pub/Sub relationships, using the SRs of the example network of Figure~\ref{fig:fog_arch}.
% These relationships are utilized, as illustrated in Figure~\ref{fig:CDN_msc}, to disseminates managed object from origin service points (\emph{i.e.} origin-oSR instance pair) to edge service points (edge-eSR instance); and, to map and deliver the requested content to end-user. 
	Recall that \emph{micro-services} are aggregate bundles of resources, named and disseminated individually under wild-card URLs; the format of which, is to be defined either by the cloud provider, or by the service provider in agreement with the cloud provider. Notably, the granularity of a resource offered by the service and the number of services in a network is a Cloud/Fog dimensioning parameter, depending on the amount of resources presented in the network.

Figure~\ref{fig:fog_ns}(b) illustrates the sequence of publication events following the name-space of Figure~\ref{fig:fog_ns}(a), when fSR$_b$ request an advertised (but not cached) resource. Following this example, service points fSR$_a$ and cSR register a subscription under the `HTTP' root scope in the PCF for requests of the FQDN it wishes to offer. This resembles the action of a server listening for requests. The Cloud data centre has a large number of resources cached, including the one requested by fSR$_b$. Therefore, cSR also subscribes to listen for requests to the micro-services it wishes to offer under the ``HTTP-Micro'' root scope. This resembles listening to requests of micro-services, issued by other service points. Here we assume that fSR$_a$ has constrained storage/compute capabilities and does not have a cached copy of the resource requested by fSR$_b$. 

When fSR$_b$ publishes a request for the FQDN, the PCF would match it with fSR$_a$ as the nearest service point. fSR$_b$ sends a request for the resource (URL) to fSR$_a$. The Fog node connected to fSR$_a$ detects that the resource is not cached; thereby, fSR$_a$ issues a publication message for a wild-card URL that encompasses the none-cached resource, under ``HTTP-Micro.'' The PCF matches cSR with fSR$_a$, finally leading to service provision from cSR to fSR$_a$. The cSR publishes the response using the same micro-service name but this time under the `HTTP' name-space. Upon reception of the requested resources, fSR$_b$ sends back to fSR$_a$ the resource requested.

\subsection*{Unicast-to-Multicast-to-Unicast}
%\label{sec:u2m2u}
Our architecture has a key advantage over existing HTTP/CoAP transmission mechanisms, that is the ability to spontaneously create multicast trees to transmit responses back from the server to synchronous, or quasi-synchronous, consumption points. This advantage is critical in providing scalable service dissemination over the network by providing an agile, flexible and dynamic way of providing multicast in the network. This stems from the ability to change the shape and size of the multicast tree, without having to adhere to pre-defined `splitting' points, and therefore, not having to address bottlenecks caused by sub-optimal tree aggregation. For a large number of users, the proposed solution allows for scalable delivery of web services using substantially less compute and network resources. 

When a group of quasi-synchronous consumption points request a service (e.g. an HTTP chunk), within a pre-defined period of time, namely the \emph{catchment interval}, the service point SR passes only the first request to the service point, while suppressing subsequent requests within the catchment interval; thereby, the service point receives a single request, for which, it issues a single response. Intuitively, this reduces the load on the processing resources by the number of requests falling within the catchment period. The service SR would then formulate a multicast group and issue a single response back, using a single multicast identifier, created following the mechanism of~\cite{Jokela2009} or~\cite{Reed2016}. The latter is used to multicast the response to the SRs of the consumption points in the group.

Notably, consumption points can be orchestrated to subscribe for services/micro-services within a predefined period; thereby, engineering the size of multicast groups according to the target savings in network resources. Next, we evaluate the proposed Fog architecture and compare it against a Fog architecture that uses DNS-redirection for mapping.

\section*{Evaluation}
%\label{sec:eva}
Here, we analyse the performance of the proposed Fog architecture in the presence of multiple Fog and Cloud service points. We quantify the savings in terms of network resources and improvements in edge-to-edge communications. This is reflected in two key-indicators: \emph{path length} and \emph{network capacity}. We model the architecture analytically over a realistic network graph from the Internet Topology Zoo~\cite{Knight2011}, namely Geant 2012 $G(V = 37, A = 116)$; and use a synthetically generated service/micro-service catalogue of $1000$ items having a Zipf popularity distribution of exponent $0.8$. Each item has a bit rate requirement in the range of $\{20, 40, 60\}$ Mbps.

We model the consumption rate by estimating the network population using the global LandScan population database~\cite{Bhaduri2007}, combined with a standard Voronoi tessellation model associating each potential end-user or end-device with the nearest network node. We assume each node to have an average load of $40\%$ of its population. The number of requests per service, in each node, is drawn from the probability of occurrence of the service.

To model service availability in the network, we assume an increasing number of service points ($\{4,8,12,16\}$) to be placed in the network. That is a combination of Cloud service points, caching a large number of web resources and Fog service points with constrained storage/compute capabilities caching only a subset of web resources. Selecting a service point has been realised with either one of two standard pseudo random algorithms: \emph{Pop}, where the nodes with largest population are the most likely chosen; or \emph{Cls}, where the nodes with the highest \emph{closeness} are most likely to be selected.

For comparison with current, DNS-based, approaches, we introduce an incremental range of DNS points $\{2,4,6,8\}$; selected by either PoP or Cls. For each combination of service points and DNS points, we run $50$ tests of randomised publications/subscriptions, and randomised placement of Fog/Cloud service points. Routing inside the core is provided through Dijkstra's shortest path algorithm, where path length is measured in hop-count.
Next, we present the collected results and elaborate on their indications.

\subsection*{Backhaul Capacity}
\begin{figure}[!t]
\includegraphics[width=1.0\linewidth]{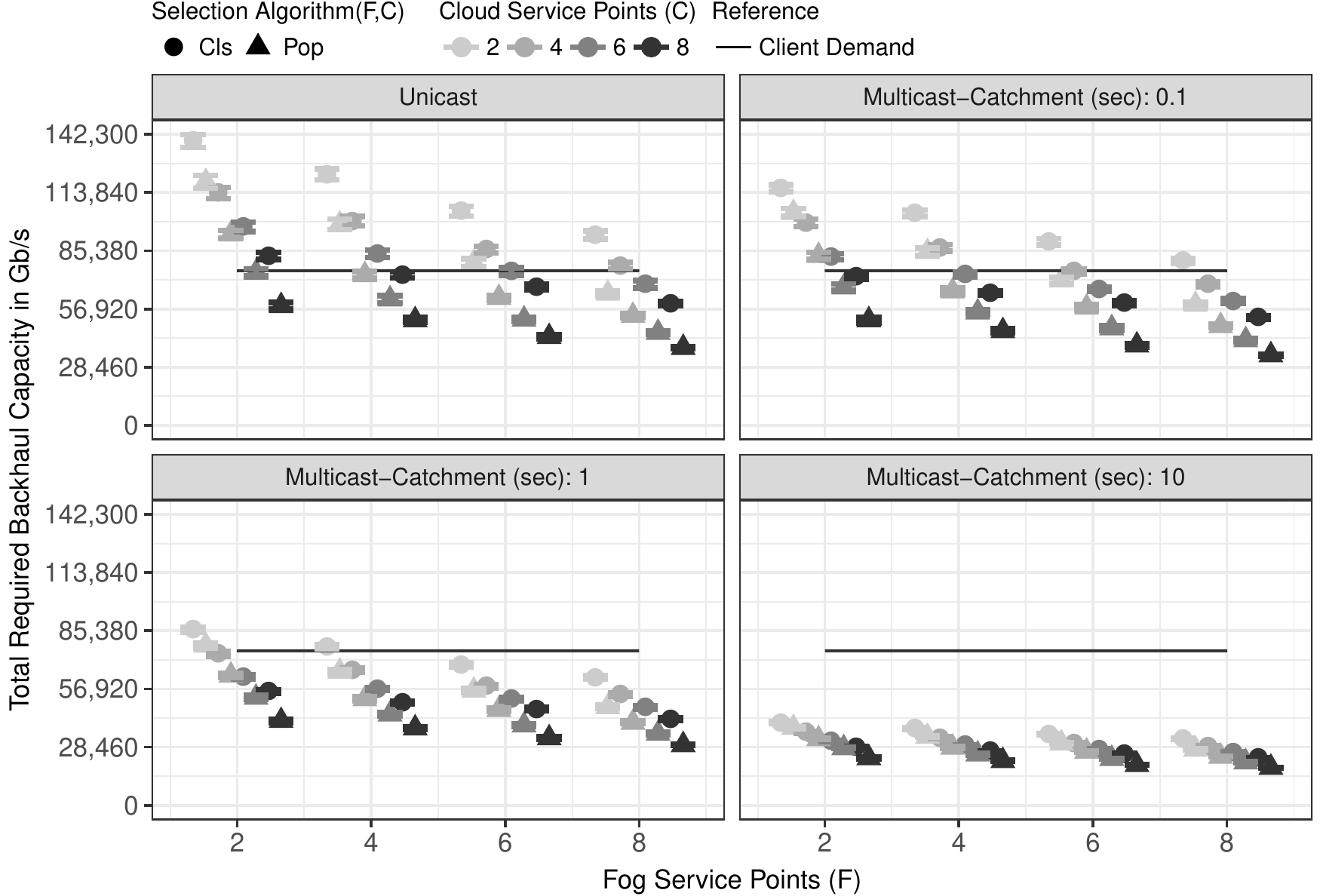}
\caption{Average required backhaul capacity in Geant network, operating the proposed Fog architecture, to satisfy a total offered demand of 70 Gb/s. The number of Cloud points is in the range $\{2,4,6,8\}$, and Fog points in the range $\{2,4,6,8\}$.}
\label{fig:icn_cap}
\end{figure}
\begin{figure}[!t]
\includegraphics[width=1.0\linewidth]{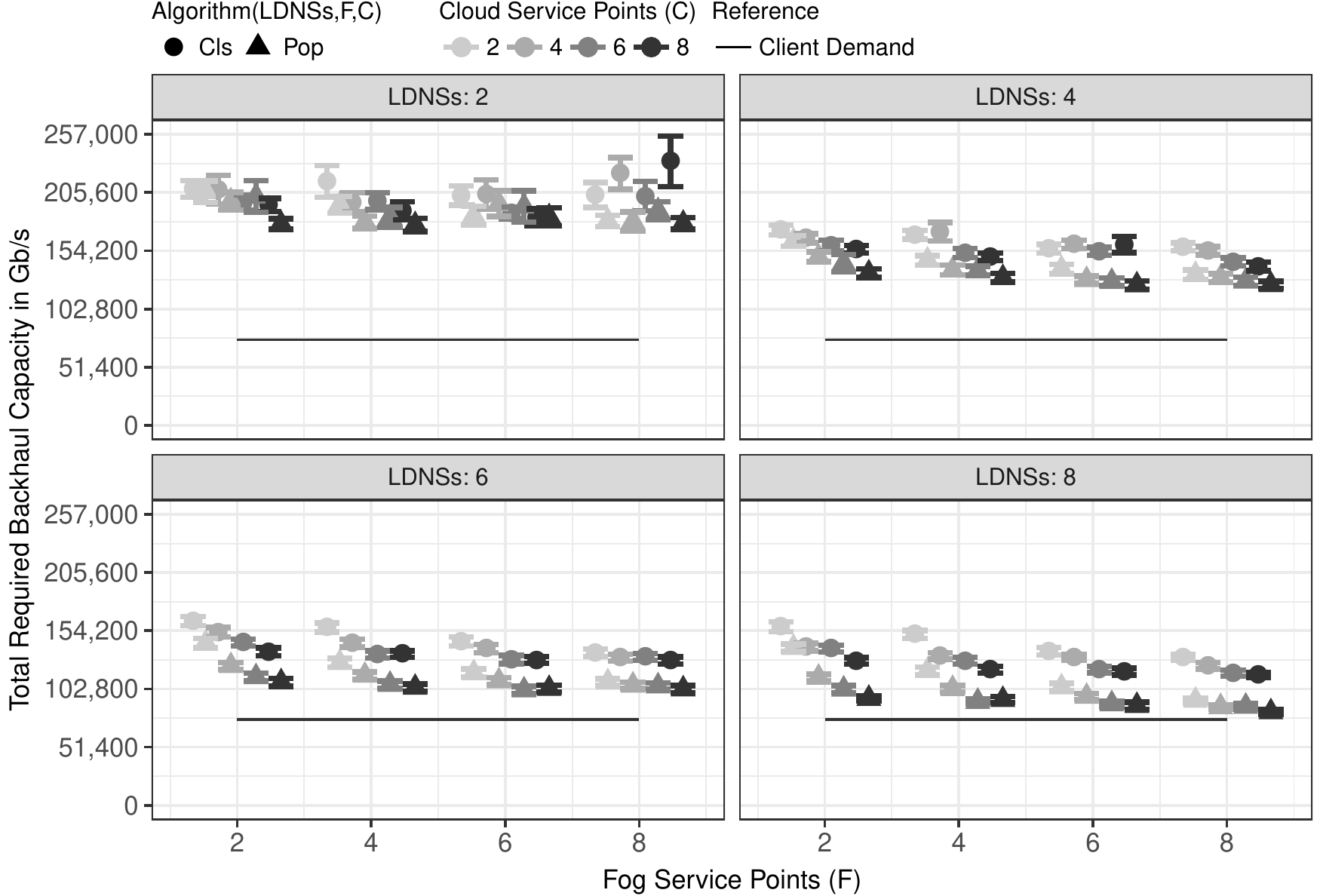}
\caption{Average required backhaul capacity in Geant network, operating a DNS-based Fog architecture, to satisfy a total offered demand of 70 Gb/s. The number of: LDNSs $\{2,4,6,8\}$, Cloud points $\{2,4,6,8\}$ and Fog points $\{2,4,6,8\}$.}
\label{fig:ip_cap}
\end{figure}
Backhaul capacity here refers to the link capacity required in the network core to allow service transmission from service to consumption points. Figure~\ref{fig:icn_cap} shows the required backhaul capacity in the proposed Fog substrate, to satisfy a total theoretical end-user consumption of $70$ Gbps. The results are shown for unicast and multicast dissemination, the latter assumes catchment intervals in the range $\{0.1, 1, 10\}$ seconds. The results indicate that the backhaul capacity is directly related to the number of Fog and Cloud service points. Increasing the number of Fog points from $2$ to $8$, for a fixed $2$ Cloud points, results in $\approx 38\%$ reduction of unicast traffic. While, increasing the number of Cloud points from $2$ to $8$, for a fixed $2$ Fog points, results in $\approx 50\%$ reduction in unicast traffic. On average, the backhaul capacity required for unicast traffic is between $50$ and $143$ Gbps, calculated by summing traffic on each link in the network. Note that the backhaul capacity is sometimes lower than the demand, because many demands are served locally from a Fog node without incurring backhaul traffic. For multicast dissemination, the bakchaul traffic is further reduced with the increase in the catchment interval. This is because a greater number of requests are falling into a single multicast group, thereby, reducing the number of groups, where each group is satisfied as a single multicast transmission.

Figure~\ref{fig:ip_cap} shows the required backhaul capacity to satisfy a total theoretical  demand of $70$ Gbps in the DNS-based Fog. The results show that the backhaul capacity is more influenced by the number of DNS points than it is by the service points (be it Cloud or Fog). For a fixed number of service points, $4$, increasing the number of local domain name servers (LDNSs) from $2$ to $8$ results in $\approx 25\%$ reduction in backhaul capacity; whilst increasing the service points from $2$ to $8$, for $8$ LDNSs, only results in $\approx 18 \%$ reduction in backhaul capacity. In all cases, the average required backhaul is approximately $1-2$ folds of the demand. This indicates that the proposed architecture has significant backhaul savings compared to a traditional DNS-based architecture.

\subsection*{Path Length}
\begin{figure}[!t]
\includegraphics[width=1.0\linewidth]{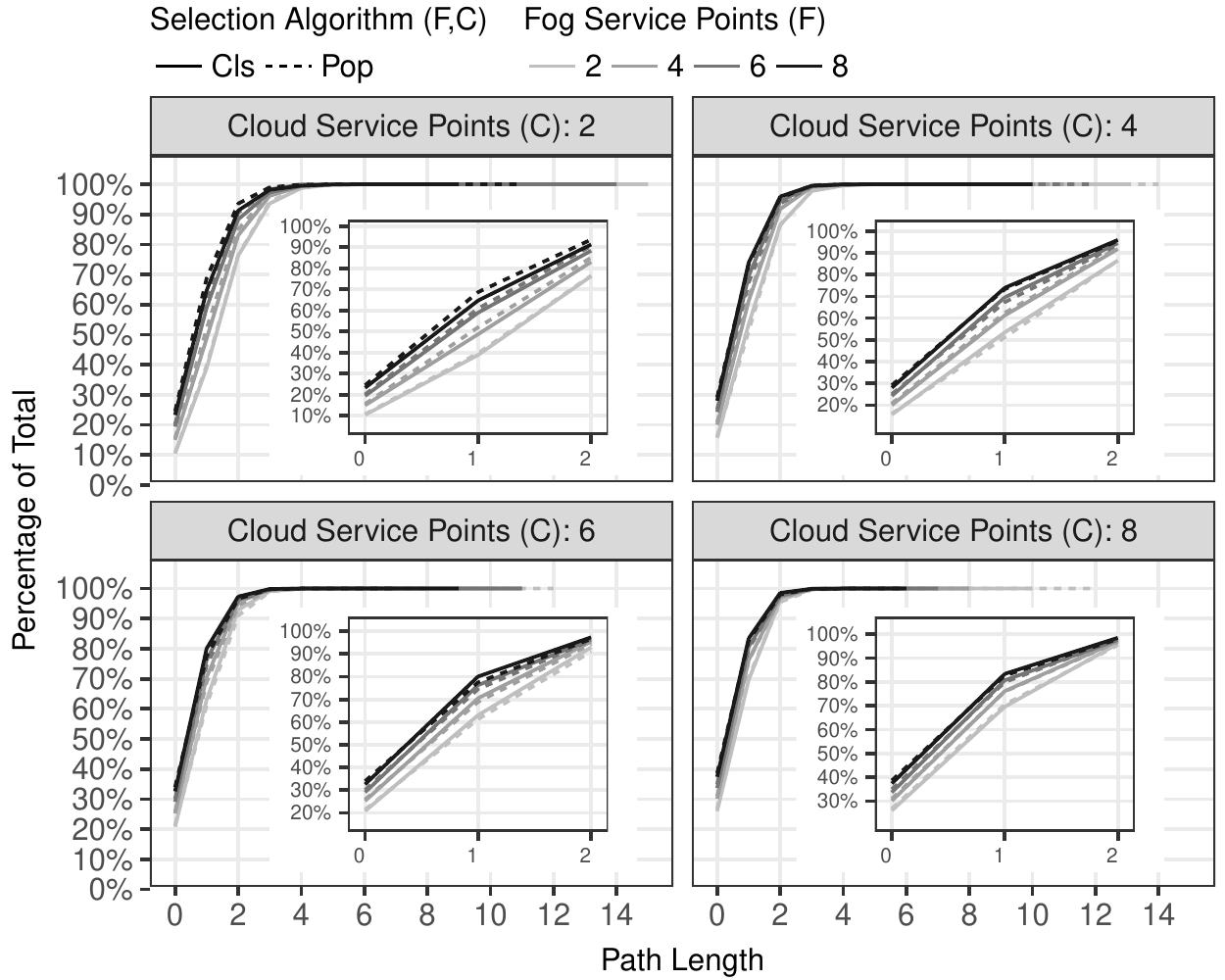}
\caption{Empirical Cumulative Distribution Function of established path lengths in the Geant network operating the proposed Fog Architecture. Fog service points are $\{2,4,6,8\}$ and Cloud points are $\{2,4,6,8\}$.}
\label{fig:icn_len}
\end{figure}
\begin{figure}[!b]
\includegraphics[width=1.0\linewidth]{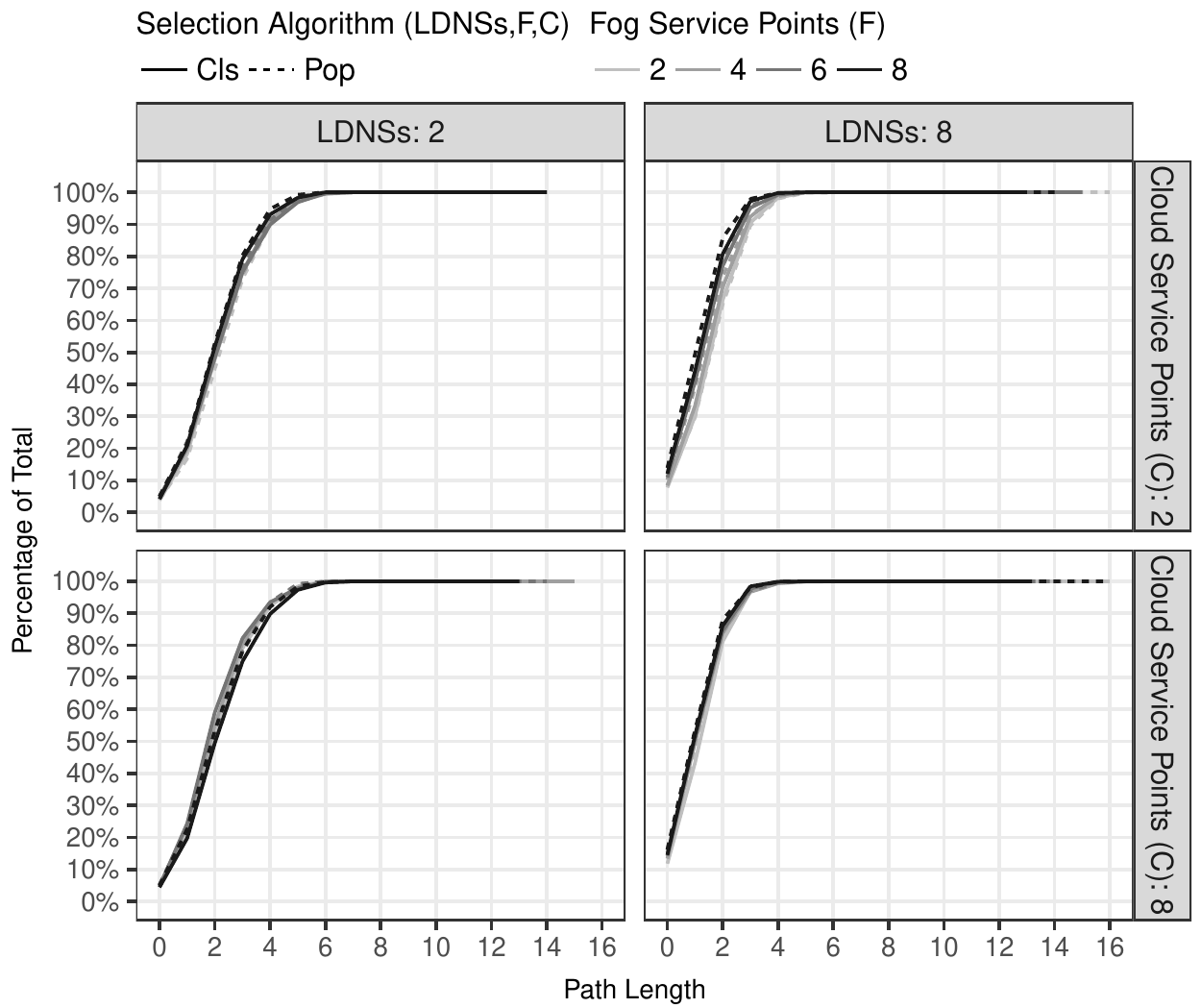}
\caption{Empirical Cumulative Distribution Function of path lengths in the Geant network operating a DNS-based Fog Architecture. Fog service points are $\{2,4,6,8\}$, Cloud point $\{2,8\}$ and LDNSs $\{2,8\}$.}
\label{fig:ip_len}
\end{figure}
Path length is used here to give an indication of relative end-to-end latency. Figure~\ref{fig:icn_len} shows the empirical cumulative distribution function (ECDF) of the paths provisioned in our architecture. When the number of service points is small, $2$, $\approx 10\%$ of paths are of $0$ length and $\approx 75\%$ are $2$ hops. The former are locally satisfied demands. Increasing the number of service points to $8$ significantly decreases the path length, with $\approx 40 \%$ localised and $\approx 97\%$ of $2$ hops. Moreover, the longest path is also reduced from $14$ hops in the first case, to $6$ hops in the second.

Figure~\ref{fig:ip_len} shows the ECDF of the paths provisioned in a DNS-based architecture. The results show considerably longer paths, affected primarily by the number of LDNSs. When LDNSs is $2$, only $5\%$ of the paths are localised and $\approx 45\%$ of $2$ hops. The longest path is $14$ hops. When LDNSs is $8$, only $15\%$ of paths are fully localised whilst $\approx 85\%$ are have $2$ hops. The longest paths are only reduced to $12$ hops.
\hfill 
 
\hfill 

\section*{Conclusion}
%\label{sec:conclusion}
Efficient Fog architectures are essential to meet the expected demands of 5/5+G and IoT. Here, we proposed a fast, fine-grained and service-based Fog architecture that avoids the inefficiencies of  DNS-based mapping which result in suboptimal service point selection. Instead, we translate edge communication into service transactions and utilise a path computation function, embraced from emerging research in ICN, to efficiently match supply with demand. The proposed architecture allows for selecting the true ``nearest'' service point. We evaluate the architecture by quantifying the required backhaul capacity and the path length, and compare it with an existing, DNS-based, architecture. Evaluation results indicate significant savings in backhaul capacity, and considerably shorter path lengths, in the proposed Fog architecture, compared to a DNS-based counterpart.

\section*{Acknowledgment}
This work was supported by the European Union funded H2020 ICT project POINT, under contract
No 643990.
This work has also been partially supported by the European Commission through project SerIoT funded by the European Union H2020 Programme under Grant Agreement No. 780139. The opinions expressed in this paper are those of the authors and do not necessarily reflect the views of the European Commission.
\newpage
\ifCLASSOPTIONcaptionsoff
  \newpage
\fi



\bibliographystyle{IEEEtran}
%\bibliography{ref}


% \end{IEEEbiography}


% % insert where needed to balance the two columns on the last page with
% %\newpage



%\vfill


\end{document}